\documentclass[10pt]{article}
\input{epsf}
\begin{document} 
\title{Mechanical Models for Higher Spin Gauge Fields}        
\author{Anders K. H. Bengtsson\footnote{e-mail: anders.bengtsson@hb.se. Work supported by the Research and Education Board at the University College of Bor{\aa}s.}}

\maketitle
\begin{center}School of Engineering\break University College of Bor{\aa}s\break All\' egatan 1, SE-50190 Bor\aa s, Sweden.\end{center}

\begin{abstract}
Returning to an old idea of a certain two-particle relativistic harmonic oscillator as an underlying mechanical model for higher spin gauge fields, various space-time pictures are discussed for the propagation and the interactions.
\end{abstract}

\maketitle
 
\section{Introduction}\label{sec:Introduction}
At the time when string theory was investigated as a theory of strong interactions, there was a contemporaneous area of research into mechanical models for quarks or partons as attempts at understanding hadrons in terms of composite systems. It went under labels such as ''infinite component wave equations'', ''bi-local field theory'' and ''dynamical groups'' (see old reviews in \cite{EighthNobelSymposium} and recent comments in \cite{Casalbuoni2006a}). Today, the correspondence between quantization of mechanical systems and field theory is of course well known and often provides useful tools for formulating both free field theories and interactions. Parts of this old work could be of relevance to the higher spin interaction problem.

Investigating this question is the motivation behind the present work. I will review one particular approach \cite{CasalbuoniLonghi1975} that indeed produced cubic higher spin vertices quite some time before the higher spin gauge interaction problem per se came into focus. Based on this work I propose an algorithm for generating $n$-vertices between massless higher spin states.

Let it also be said that this is very much work in progress and it is described as it was at the time of the 4-th EU RTN Workshop in Varna, 11 - 17 September 2008. I plan to return to this topic and treat it more thoroughly elsewhere.

\section{Constraints}\label{sec:Constraints}
We think of a mechanical system consisting of two point particles with coordinates $t^\mu$ and $b^\mu$ or equivalently center of mass $x^\mu$ and relative coordinates $\xi^\mu$ defined by
\begin{equation}\label{eq:TopBottomTranscription}
x^\mu={1\over 2}(t^\mu+b^\mu),\quad\xi^\mu={1\over 2}(t^\mu-b^\mu).
\end{equation}

Canonical momenta are $u_\mu$, $d_\mu$ or center of mass and relative $p_\mu$, $\pi_\mu$ with
\begin{equation}\label{eq:UpDownTranscription}
p^\mu=u^\mu+d^\mu,\quad\pi^\mu=u^\mu-d^\mu.
\end{equation}

For the phase space $(\xi,\pi)$ we have the oscillator transcription (dimension carried by the parameter $\kappa$)
\begin{equation}\label{eq:OscillatorTranscription2}
\alpha_\mu={1\over\sqrt 2}(\kappa\pi_\mu-{i\over\kappa}\xi_\mu),\quad\alpha_\mu^\dagger={1\over\sqrt 2}(\kappa\pi_\mu+{i\over\kappa}\xi_\mu).
\end{equation}

In terms of these phase space variables we know what the correct first class constraints should be in order to arrive at an infinite tower of higher spin gauge fields. The constraints are
\begin{equation}\label{eq:HSConstraints}
G_0={1\over 2}p^2,\quad G_+=\alpha\cdot p,\quad G_-=\alpha^\dagger\cdot p,
\end{equation}

spanning the simple algebra
\begin{equation}\label{eq:AlgebraHSConstraints}
[G_+,G_-]=2G_0,
\end{equation}

with all other brackets zero.

The BRST charge Q that results from this theory is of course precisely the one that produces an infinite tower of gauge fields with actions unified in $\langle\Phi|Q|\Phi\rangle$ \cite{SiegelZwiebach1987,OuvryStern1987a,AKHB1987a,Meurice1988} \footnote{This basic scheme has been independently rediscovered several times since the mid 1980's and has subsequently been extended in various directions by several groups of authors. For a recent review of this modern literature, see \cite{FotopoulosTsulaia2008a}}.

We can consider the $\langle\Phi|Q|\Phi\rangle$ setup as simply a practical mnemonic for deriving the Fronsdal free field equations. If, however, we want to base some kind of space-time picture for the interactions, it would seem that we need an underlying space-time mechanical model.

The general constraint structure of models of this type is simple but can be organized in several different ways. Excluding explicit $x$, there are three more bilinear constraints to consider
\begin{equation}
T={1\over 2}\alpha\cdot\alpha,\quad T^\dagger={1\over 2}\alpha^\dagger\cdot\alpha^\dagger,\quad N={1\over 2}(\alpha\cdot\alpha^\dagger+\alpha^\dagger\cdot\alpha).
\end{equation}

Quantum ordering ambiguities can only arise for $N$.

By choosing different subsets of these constraints, and different linear combinations, we can study various types of bi-local models.

\paragraph{Case I. Reducible tower of higher spin gauge fields}
Basing a $\langle\Phi|Q|\Phi\rangle$ theory on just the three first class constraints $G_0,G_-,G_+$ yields an infinite tower of higher spin gauge fields. It is reducible in the sense that we are not removing the double traces $\varphi^{\prime\prime}$ of the gauge fields and the corresponding traces of the gauge parameters $\xi^\prime$. Effectively this means that we allow the propagation of lower spin fields. The details was analysed in \cite{AKHB2007a}.

\paragraph{Case II. Irreducible tower of higher spin gauge fields}
If we supply the theory with the constraints $T$ and $T^\dagger$, but not with $N$, these two become second class. When augmented with ghost contributions and applied to the states $|\Phi\rangle$ they yield the double trace and trace constraints and we precisely reproduce the Fronsdal equations.

\paragraph{Case III. Irreducible single higher spin gauge field}
If we add $N-\lambda$ with a constant non-negative integer $\lambda$ as a constraint, this together with $T$ and $T^\dagger$ become second class and we effectively describe one single higher spin gauge field with a definite helicity $\lambda$. This system is very interesting. It was investigated by Casalbuoni, Dominici and Longhi \cite{CasalbuoniLonghi1975,CasalbuoniDominiciLonghi1976} in the mid 1970's. It results from the analysis of a two-particle Lagrangian describing a specific rigid motion of a string. As we will see in the next section, it provides us with an algorithm for interaction vertices for higher spin states.

\paragraph{Case IV. Reducible single higher spin gauge field}
Another option is to take $G_0,G_-,G_+$ and $N-\lambda$ as constraints but not $T, T^\dagger$. These four constraints then become first class. We fix the spin to $\lambda$ but we get extra propagating lower spin components as in case I.

\paragraph{Case V. Regge trajectory of massive higher spin fields}
Returning to the first case and combining $G_0$ and $N$ into one constraint $G_0+N$, all the rest of the constraints become second class. We get a Regge trajectory of massive higher spin fields. Gauge invariance is of course lost since these transformations are otherwise generated by the $G_-$ and $G_+$ constraints. A system of this kind was analysed in \cite{GershunPashnev1988}. It results from discretising the bosonic string into two end-point particles. Discretising into more than two particles (corresponding to more than one oscillator) yield several trajectories.

\section{A mechanical model}
It is indeed quite remarkable that such an extremely simple constraint structures as in case I and II above is sufficient to reproduce all of the Fronsdal spin $s$ actions. This simplicity is however somewhat deceptive when we try to extend the theory to include interactions. We simply don't have enough structure to support interactions. The only guiding principle is to extend gauge invariance to all orders in the fields. This tells us that the gauge algebra is a strongly homotopy Lie algebra, but not much more \cite{FulpLadaStasheff2002,AKHB2005a,Vasiliev2005a}.

Turning then to cases III and IV, where the essential point is that $N-\lambda$ is a constraint, we fix the spin to a certain value $\lambda$. The actions and field equations stay the same. Physically interpreted, the model now describes physical oscillators, not just practical mnemonics. This extra structure allows us to write down an algorithm that generates higher spin vertices.

The model of \cite{CasalbuoniLonghi1975} corresponds to a rigid motion of the string (also called ''straight line string'' in the literature) with Lagrangian in terms of center of mass $x$ and relative $\xi$ coordinates
\begin{equation}\label{eq:Lagrangian}
L={1\over 2\kappa^2}\Big(\sqrt{((\dot x-\dot \xi)\cdot\xi)^2-(\dot x-\dot \xi)^2{\xi}^2}+(\xi\rightarrow -\xi)\Big).
\end{equation}

A careful Dirac analysis of this action reveals two branches of linear combinations of the constraints, one corresponding to massless states and one corresponding to massive states (recently investigated in \cite{KamimuraShiseki2008a}). This makes physical sense. The relative coordinate oscillations yield massless states, while the rotations yield massive states. Here we will focus on the massless states. The massive branch is equivalent to the two-particle discretised string of \cite{GershunPashnev1988} (Case V above).

We are then interested in harmonic oscillator states $|\mu_1,\ldots,\mu_n\rangle=\alpha_{\mu_1}^\dagger\cdots\alpha_{\mu_n}^\dagger|0\rangle$ which we write for short as $|\mu_{(n)}\rangle$. In BRST-Lagrangian higher spin theory these states are used as a basis for classical higher spin fields $\phi^{\mu_1\cdots\mu_n}(x)$.

The configuration space states corresponding to the Fock space states are
\begin{equation}\label{eq:ConfigStates1}
f_{\mu_1\ldots\mu_n}(\xi)=\langle\xi|\mu_1,\ldots,\mu_n\rangle.
\end{equation}

Generalising the generating function $f(\xi,J)=\exp[(-J^2+2J\xi)/\kappa^2]$ for one-dimensional non-relativistic oscillator states and inserting a factor for the ground state wave function $\langle\xi|0\rangle=\exp[-\xi^2/2\kappa^2]$ and a suitable normalisation $c_n$, the configuration space states are given by partial derivatives with respect to the sources $J^{\mu_i}$
\begin{equation}\label{eq:ConfigStatesGeneratingFormula}
f_{\mu_1\ldots\mu_n}(\xi)=c_n\exp[-\xi^2/2\kappa^2]{\partial^{(n)}\over \partial^{\mu_1}\ldots\partial^{\mu_n}}\exp[(-J^2+2J\cdot\xi)/\kappa^2]\Big|_{J=0}.
\end{equation}

The functions $f_{\mu_1\ldots\mu_n}(\xi)$ serve equally well as a basis for the higher spin gauge fields. Indeed we can write
\begin{equation}
\Phi(x,\xi)=\langle\xi|\Phi(x,\alpha^\dagger)\rangle=\langle\xi|\sum_{n=0}^\infty\phi^{\mu_1\ldots\mu_n}(x)|\mu_1,\ldots,\mu_n\rangle=\sum_{n=0}^\infty\phi^{\mu_1\ldots\mu_n}(x)f_{\mu_1\ldots\mu_n}(\xi).
\end{equation}

\section{Cubic interactions}\label{sec:CubicInteractions}
Let us review the Casalbuoni-Dominici-Longhi cubic vertex and then generalise to higher order vertices.

Consider then two bi-particle systems like the ones described above colliding to produce a third. It is natural to consider end-point collisions, for instance a $t$ colliding with a $b$. Consider therefore the picture describing the three-vertex. We denote the end-point coordinates with $(t_i,b_i)$ with $i\in\{1,2,3\}$ and the collision points with $x_1,x_2,x_3$ all taken at a common time $\tau$. Conventionally we consider all the states participating in the collision as being incoming. There is a natural cyclic symmetry on the integers labeling the incoming states. It is obvious how to generalise to an $n$-vertex.

\begin{figure}[h]
\epsfbox{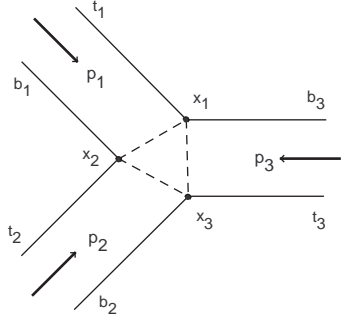} \caption{Spacetime picture of cubic
vertex.} \label{fig:CubicVertex}
\end{figure}

We work with incoming states $\langle t,b|$ of definite momentum and spin represented by $\langle p;\mu_{(n)}|$. The model is gauge fixed so there are no mechanical ghost degrees of freedom (this restriction eventually have to be lifted). The natural overlap conditions are coordinate continuity, which in the notation of the figure amounts to
\begin{eqnarray}\label{eq:CoordinateContinuity}
b_1=x_2=t_2,\\\nonumber
b_2=x_3=t_3,\\\nonumber
b_3=x_1=t_1.\\
\end{eqnarray}

The first quantised vertex operator is (up to numerical and dimensional factors)
\begin{equation}\label{eq:CubicVertexOperator}
|V(\tau)\rangle\sim g\int d^4x_1d^4x_2d^4x_3|x_1,x_2;\tau\rangle|x_2,x_3;\tau\rangle|x_3,x_1;\tau\rangle.
\end{equation}

Computing the matrix element with the incoming states $\langle p_i;\mu_{i(n)}|$ yields
\begin{equation}
V_{123}\sim g\int d^4x_1d^4x_2d^4x_3\langle p_1;\mu_{1(n_1)}|x_1,x_2;\tau\rangle\langle p_2;\mu_{2(n_2)}|x_2,x_3;\tau\rangle\langle p_3;\mu_{3(n_3)}|x_3,x_1;\tau\rangle
\end{equation}

Next using the overlap conditions we get (and similarly for the $i=2,3$)
\begin{equation}
\langle p_1;\mu_{1(n_1)}|x_1,x_2;\tau\rangle=\exp[-ip_1\cdot x_{12}]f(z_{12}),
\end{equation}

with
\begin{equation}
x_{12}={1\over 2}(x_1+x_2),\quad z_{12}={1\over 2}(x_1-x_2).
\end{equation}

The vertex function thus become
\begin{eqnarray}\label{(eq:VertexFunction}
V_{123}=V(p_i,J_i)\sim g\int d^4x_1d^4x_2d^4x_3\exp[-i(p_1\cdot x_{12}+p_2\cdot x_{23}+p_3\cdot x_{31})]\times\\\nonumber
\exp\big[{1\over\kappa^2}[-{1\over2}(z_{12}^2+z_{23}^2+z_{31}^2)-(J_1^2+J_2^2+J_3^3)+2(J_1\cdot z_{12}+J_2\cdot z_{23}+J_3\cdot z_{31})]\big].
\end{eqnarray}

Performing the integrations we get up to numerical and dimensional factors
\begin{eqnarray}
V(p_i,J_i)\sim g\exp\Big[{1\over3\kappa^2}(J_1^2+J_2^2+J_3^3)-{4\over3\kappa^2}(J_1\cdot J_2+J_2\cdot J_3+J_3\cdot J_1)+\\\nonumber
{2i\over3}\big[J_1\cdot(p_2-p_3)+J_2\cdot(p_3-p_1)+J_3\cdot(p_1-p_2)\big]-{\kappa^2\over6}(p_1^2+p_2^2+p_3^2)\Big].
\end{eqnarray}

As noted, the $J$'s can be thought of as sources for the higher spin states. So performing partial derivatives with respect to various combinations of $J$'s, and then setting them to zero, we get expressions for particular spin states interacting. For instance, computing
\begin{equation}\label{eq:YMCubic1}
{\partial\over\partial J_1^\mu}{\partial\over\partial J_2^\nu}{\partial\over\partial J_3^\rho}V(p_i,J_i)\Big|_{J_i=0}
\end{equation}

results in the Yang-Mills vertex $\sim gf_{abc}\big(\eta_{\mu\nu}(p_1-p_2)_\rho+\eta_{\nu\rho}(p_2-p_3)_\mu+\eta_{\rho\mu}(p_3-p_1)_\nu\big)$.

Now it is quite clear that this scheme can be generalised to an algorithm for computing ''candidate'' interactions to any order and any spin. I say ''candidate'' since it remains to show consistency of the resulting interactions.

\section{Vertex algorithm}
Let us outline the algorithm for arbitrary vertex order $n$ in a simplified notation. Thus let $p$ and $J$ denote a set of $n$ incoming momenta $p_i$ and corresponding sources $J_i$. We write the vertex generating function $V(p,J)$ as $V(p,J)=\kappa^{2(n-2)}\exp[\Delta]$ where
\begin{equation}\label{eq:VertexDelta}
\Delta={1\over\kappa^2}N_{ij}J_iJ_j+M_{ij}J_ip_j+\kappa^2 p^2,
\end{equation}

and where the indices $i,j$ include also space-time indices. Formally we sum from 1 to $n$ but we keep in mind that $N$ and $M$ are really combinations of the metric $\eta_{\mu\nu}$ and Kronecker $\delta_{ij}$ depending on the vertex order. The precise form of $V$ follows from generalising the construction for cubic interaction in the previous section to an $n$-vertex and performing the integrations over the $n$ end-point overlap coordinates $x_i$. This will result in a generic expression of the form (\ref{eq:VertexDelta}). Thus, the vertex order $n$ is implicit and will show up in the actual form of the $N$'s and $M$'s.

Vertices for various combinations of fields are extracted from $V(p,J)$ by performing appropriate partial derivatives and setting the sources to zero. The following formulas are useful (numerical factors are ignored)
\begin{eqnarray}\label{eq:Formulas1}
{\partial\over\partial J_i}\Delta&\sim&{1\over\kappa^2}N_{ij}J_j+M_{ij}p_j\equiv\varphi_i\rightarrow Q_i=M_{ij}p_j,\\\nonumber
{\partial^2\over\partial J_i\partial J_j}\Delta&\sim&{1\over\kappa^2}N_{ij}\rightarrow{1\over\kappa^2}N_{ij},\\\nonumber
{\partial^3\over\partial J_i\partial J_j\partial J_k}\Delta&=&0.
\end{eqnarray}

Putting this scheme to work for the spin 1 cubic vertex we get
\begin{equation}
{\partial^3\over\partial J_1\partial J_2\partial J_3}\kappa^2\exp[\Delta]\Big|_{J_i=0}=(N_{12}Q_3+N_{23}Q_1+N_{31}Q_2+\kappa^2 Q_1Q_2Q_3)\exp(\kappa^2 p^2).
\end{equation}

Remembering that the $N$'s are combinations of spacetime metrics $\eta$ and the $Q$'s of spacetime metrics $\eta$ and momentum factors $p$, we see that we reproduce the Yang-Mills cubic vertex when the $NQ$-terms are expanded. The ${\cal O}(\kappa^2)$ terms with combinations of three momentum factors is expected to occur \cite{AKHB2007a}. Such higher derivative interactions (in this case for spin 1) has been found \cite{BekaertBoulangerCnockaert2006a} and corresponds to the spin 1 fields that accompany the spin 3 fields in cases where the tracelessness constraints aren't imposed (Case I above).

In this way we can compute arbitrary order vertices for arbitrary combinations of massless fields and it is clear that we get the correct overall momentum powers.

Much remain to be done however. The vertices are covariant but field theoretically they correspond to a definite choice of field redefinitions. We have to understand how to check gauge invariance to all orders, something which presumable requires introducing mechanical ghosts and the concomitant BRST-machinery. A crucial calculational test is to compare to the spin 3 cubic vertex of \cite{BerendsBurgersvanDam1984}.

This, scheme, if it can be shown to be consistent with gauge invariance to all orders, holds out the hope of providing a systematic method of obtaining higher spin vertices. Some fairly simple organizing structure to the interactions is precisely what is lacking. I hope to return to these issues elsewhere.

\section*{acknowledgement}
I would like to thank Bo Sundborg for discussions we had almost twenty years ago, the notes of which I found useful to return to when writing up the present work. Unfortunately the ideas didn't quite jell at that time. Too many pieces of the puzzle were still missing.


\begin{thebibliography}{10}

\bibitem{EighthNobelSymposium}
N.~Svartholm, editor.
\newblock {\em Elementary Particle Theory. Relativistic Groups and
  Analyticity}. Almqvist-Wiksell, Wiley Interscience Division, 1968.

\bibitem{Casalbuoni2006a}
R.~Casalbuoni.
\newblock Majorana and the infinite component wave equations.
\newblock 2006.
\newblock hep-th/0610252.

\bibitem{CasalbuoniLonghi1975}
R.~Casalbuoni and G.~Longhi.
\newblock A geometrical model for nonhadrons and its implications for hadrons.
\newblock {\em Nouvo Cimento}, 25 A:482--502, 1975.

\bibitem{SiegelZwiebach1987}
W.~Siegel and B.~Zwiebach.
\newblock Gauge string fields from the light-cone.
\newblock {\em Nucl. Phys. B}, 282:125--141, 1987.

\bibitem{OuvryStern1987a}
S.~Ouvry and J.~Stern.
\newblock Gauge fields of any spin and symmetry.
\newblock {\em Phys. Lett. B}, 177:335--334, 1987.

\bibitem{AKHB1987a}
A.~K.~H. Bengtsson.
\newblock A unified action for higher spin gauge bosons from covariant string
  theory.
\newblock {\em Phys. Lett. B}, 182:321--325, 1987.

\bibitem{Meurice1988}
Y.~Meurice.
\newblock From points to gauge fields.
\newblock {\em Phys. Lett. B}, 186:189--194, 1988.

\bibitem{FotopoulosTsulaia2008a}
A.~Fotopoulos and M.~Tsulaia.
\newblock {Gauge Invariant Lagrangians for Free and Interacting Higher Spin
  Fields. A Review of the {BRST} formulation}.
\newblock 2008.

\bibitem{AKHB2007a}
A.~K.~H. Bengtsson.
\newblock Structure of higher spin gauge interactions.
\newblock {\em J. Math. Phys.}, 48:072302, 2007.
\newblock hep-th/0611067.

\bibitem{CasalbuoniDominiciLonghi1976}
D.~Dominici R.~Casalbuoni and G.~Longhi.
\newblock On the second quanitization of a composite model for nonhadrons.
\newblock {\em Nouvo Cimento}, 32 A:265--275, 1976.

\bibitem{GershunPashnev1988}
V.~D. Gershun and A.~I. Pashnev.
\newblock Relativistic system of interacting points as a discrete string.
\newblock {\em Theor. Math. Phys.}, 73:294--301, 1987.

\bibitem{FulpLadaStasheff2002}
R.~Fulp, T.~Lada, and J.~Stasheff.
\newblock {Sh-Lie} algebras induced by gauge transformations.
\newblock {\em Commun. Math. Phys.}, 231:25--43, 2002.
\newblock math.QA/0012106.

\bibitem{AKHB2005a}
A.~K.~H. Bengtsson.
\newblock An abstract interface to higher spin gauge field theory.
\newblock {\em J. Math. Phys.}, 46:042312, 2005.
\newblock hep-th/0403267.

\bibitem{Vasiliev2005a}
M.A. Vasiliev.
\newblock Actions, charges and off-shell fields in the unfolded dynamics
  approach.
\newblock {\em Int.J. Geom. Meth. Mod. Phys.}, 3:37--80, 2006.
\newblock hep-th/0504090.

\bibitem{KamimuraShiseki2008a}
K.~Kamimura and D.~Shiseki.
\newblock Massive rigid string model and its supersymmetric extension.
\newblock {\em Nucl. Phys. B}, 806:489, 2009.

\bibitem{BekaertBoulangerCnockaert2006a}
N.~Boulanger X.~Bekaert and S.~Cnockaert.
\newblock Spin three gauge theory revisited.
\newblock {\em JHEP}, 0601:052, 2006.
\newblock hep-th/0508048.

\bibitem{BerendsBurgersvanDam1984}
F.~A. Berends, G.~J.~H. Burgers, and H.~van Dam.
\newblock On spin three self interactions.
\newblock {\em Z. Phys. C}, 24:247--254, 1984.

\end{thebibliography}
\end{document}